\documentclass[10pt,conference,letterpaper]{IEEEtran}
\IEEEoverridecommandlockouts
\pdfoutput=1

\usepackage{cite}
\usepackage{caption}
\usepackage{amsmath,amssymb,amsfonts}
\usepackage[ruled,vlined]{algorithm2e}
\usepackage{algpseudocode}
\usepackage{graphicx}
\usepackage{textcomp}
\usepackage{xcolor}
\def\BibTeX{{\rm B\kern-.05em{\sc i\kern-.025em b}\kern-.08em T\kern-.1667em\lower.7ex\hbox{E}\kern-.125emX}}

\usepackage{hyperref}
\hypersetup{
    colorlinks = true,
    citecolor = magenta,
    linkcolor = purple
}
      
\usepackage{xcolor}
\usepackage{tabularx, makecell, linegoal}

    \usepackage{ifthen}
    
\newboolean{ShowComments}
\setboolean{ShowComments}{true}  
\ifthenelse{\boolean{ShowComments}}%
	{
		\newcommand{\ColorComment}[3]{%
				{\colorbox{#1}{\color{white}   \textsf{\textbf{#2}}} \textcolor{#1}{#3}}}

	}%
	{
		\newcommand{\ColorComment}[3]{}

	}%

\newcommand{\OPL}{\textsc{OPL}\xspace}
\definecolor{rdvcolor}{rgb}{0,0.5,0}
\definecolor{satohcolor}{RGB}{254,0,0}
\definecolor{michalcolor}{RGB}{255,127,80}
\definecolor{naphanncolor}{RGB}{112, 51, 173}
\definecolor{shotacolor}{RGB}{0, 0, 255}
\definecolor{commentcolor}{RGB}{160, 60, 160}
\definecolor{generalcolor}{RGB}{130, 70, 178}

\begin{document}

\title{Automatic Configuration Protocols for Optical Quantum Networks

\thanks{This work was supported by JST [Moonshot R\&D Program] Grant Number [JPMJMS226C].}
}

\author{
\IEEEauthorblockN{
    Amin Taherkhani\IEEEauthorrefmark{1},
    Andrew Todd\IEEEauthorrefmark{1},
    Kentaro Teramoto\IEEEauthorrefmark{2},     
    Rodney Van Meter\IEEEauthorrefmark{4}\IEEEauthorrefmark{3}, Shota Nagayama\IEEEauthorrefmark{3}\IEEEauthorrefmark{2},
}\\

\IEEEauthorblockA{\IEEEauthorrefmark{1}\textit{Graduate School of Media and Governance, Keio University Shonan Fujisawa Campus, Kanagawa, Japan}}
\IEEEauthorblockA{\IEEEauthorrefmark{2}\textit{mercari R4D, Mercari, Inc., Tokyo, Japan}}
\IEEEauthorblockA{\IEEEauthorrefmark{3}\textit{Quantum Computing Center, Keio University, Kanagawa, Japan}}
\IEEEauthorblockA{\IEEEauthorrefmark{4}\textit{Faculty of Environment and Information Studies, Keio University Shonan Fujisawa Campus, Kanagawa, Japan}}
\IEEEauthorblockA{\IEEEauthorrefmark{5}\textit{Graduate School of Media Design, Keio University Hiyoshi Campus, Yokohama, Kanagawa, Japan}}

\{amin, rdv\}@sfc.wide.ad.jp, at@auspicacious.org, zigen@mercari.com,
shota@qitf.org}


\maketitle
\begin{abstract}
Before quantum networks can scale up to practical sizes, there are many deployment and configuration tasks that must be automated. Currently, quantum networking testbeds are largely manually configured: network nodes are constructed out of a combination of free-space and fiber optics before being connected to shared single-photon detectors, time-to-digital converters, and optical switches. Information about these connections must be tracked manually; mislabeling may result in experimental failure and protracted debugging sessions. In this paper, we propose protocols and algorithms to automate two such manual processes. First, we address the problem of automatically identifying connections between quantum network nodes and time-to-digital converters. Then, we turn to the more complex challenge of identifying the nodes attached to a quantum network's optical switches. Implementation of these protocols will help enable the development of other protocols necessary for quantum networks, such as network topology discovery, link quality monitoring, resource naming, and routing. We intend for this paper to serve as a roadmap for near-term implementation.
\end{abstract}


\section{Introduction}
The planning and deployment of quantum system-area optical networks~\cite{rdv2014quantum}, whether to provide interconnection infrastructure for quantum multicomputer architectures~\cite{rdv2006architecture,BARRAL2025100747} or quantum cloud services for blind computation~\cite{broadbent2009universal}, requires addressing many design decisions, ranging from topologies and architecture to protocols~\cite{Monroe2014modular,sakuma2024optical,10.17487/RFC9340}. 

Meanwhile, considerable efforts are being made to prototype small-scale quantum networks in academic and industry testbeds around the world~\cite{pompili2021realization,Alshowkan22,Krutyanskiy2023ion,QUANTNET2024,amin2024}. However, experimental demonstration of useful quantum networks will require scaling up the number of connected nodes far beyond current prototypes. Just as in classical network engineering, larger scale implies a need for more automation. Today's manually configured testbeds cannot serve the needs of increasingly complex experiments.

To expand on this point, the main focus of quantum network experiments today is the analysis and benchmarking of network links based on typical performance indicators such as fidelity, entanglement generation rate, and entanglement distribution rate. Automation is focused on monitoring these characteristics \cite{eisert2020quantum}. However, such experiments often assume that automatic determination of the composition and topology of these network links is trivial. In reality, the one-way nature of testbed lightpaths and the difficulty of encoding and receiving classical information on these lightpaths during quantum system operation create significant challenges for the automated discovery of physical network topology in all but the simplest testbed configurations.

Solving the problem of automated network topology discovery will require the creation of tools for every layer of hardware comprising the network. In this paper, we start from the bottom up and propose two low-level system configuration detection protocols. We intend to implement these protocols in a quantum network testbed.

%
\begin{figure*}[t]
    \centering
    \includegraphics[width=0.9\linewidth]{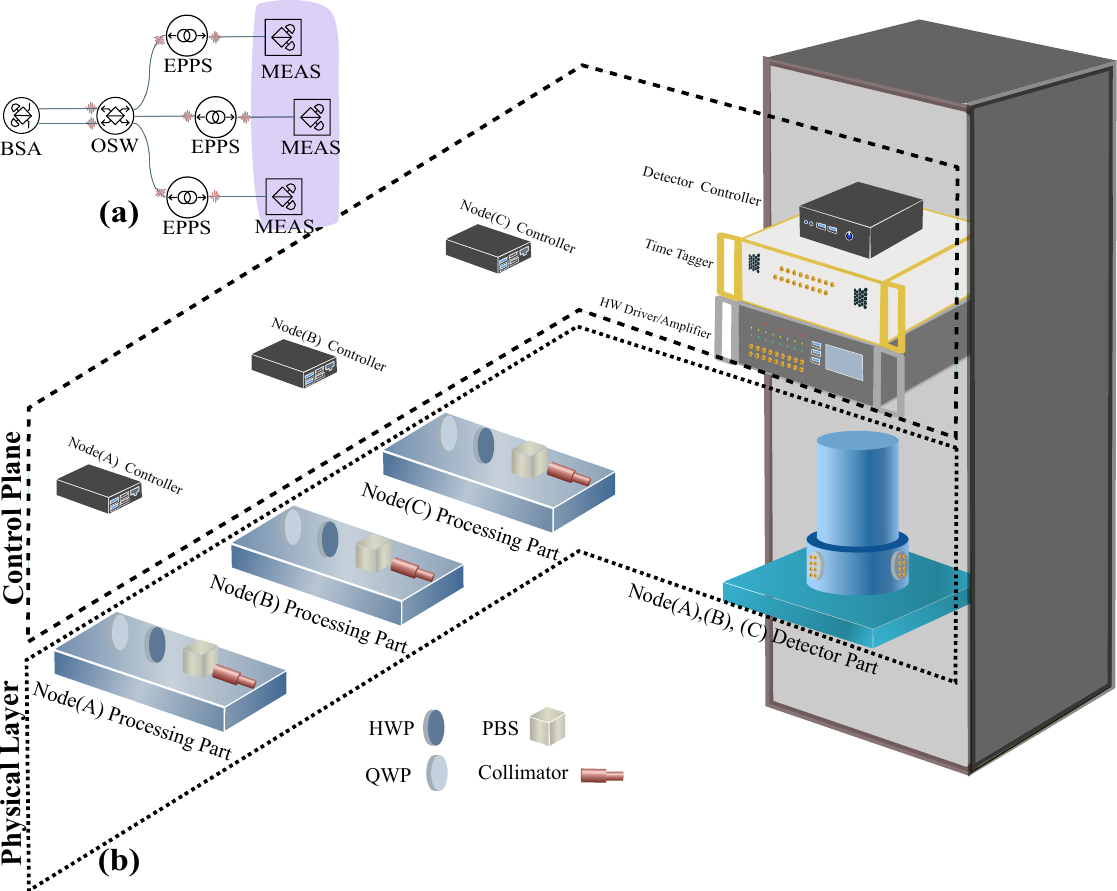}
    \caption{(a) A high level logical quantum network diagram in a photonic synchronization domain (PSD) showing MEAS, EPPS, OSW, and BSA nodes. (b) A sample real implementation of a subset of the network (the MEAS nodes in the highlighted block) with optical devices in the physical plane and node controllers in the classical control plane. The detector components of the nodes are physically separated from the processing part of the nodes, necessitating some means of logical association. Furthermore, the node controllers in classical control plane lack information about the physical connectivity and neighboring relationship of the optical channels, requiring explicit inter-node channel identification.}
    \label{fig:detector_channels_match}
\end{figure*}


\subsection{Toward independent network nodes}

From a logical perspective, testbed networks are comprised of network nodes. As identified by Van Meter et al.\cite{van2022quantum}, some of the most common nodes include:

\begin{itemize}
    \item Measurement (MEAS): Measures photons in at least one basis. As such, it will have at least one SNSPD. Often used as a stand-in for full quantum computer or quantum memory nodes during testing.
    \item Bell state analyzer (BSA): Used to perform entanglement swapping. Requires at least two SNSPDs.
    \item Entangled photon pair source (EPPS): Used to produce pairs of entangled photons. One half of the pair may travel to a MEAS node while the other half is routed to a BSA.
    \item Optical switch (OSW): Used to create configurable lightpaths between light sources such as EPPS nodes and light sinks such as MEAS and BSA nodes. In current implementations, OSW nodes tend to have very basic functionality: they have no way to measure the light passing through them.
\end{itemize}

Readers accustomed to thinking about classical networking hardware might imagine these nodes as looking like network cards and routers, each node cleanly separated from all others. In the future this may be the case. However, the current reality is much more complex: due to cost and space considerations, hardware is often shared between nodes. Thus, setting up an experimental testbed is considerably more complex than simply plugging network nodes into each other.

This paper examines two common real-world configuration problems and proposes solutions to help bridge this gap between current, piecemeal systems and future, integrated systems. 

\subsubsection{Automatic node-to-TDC configuration}

MEAS and BSA nodes integrate single-photon detectors, most often superconducting nanowire single-photon detectors (SNSPDs). SNSPDs themselves have only the most basic means of output: for each photon detected, they emit an electrical pulse. The electrical pulse is then time-tagged by a high precision time-to-digital converter (TDC). An ordinary classical computer can then read the TDC's time-tag data and analyze it.

SNSPDs must be located in extremely low-temperature cryostats. Moreover, most TDC applications, such as coincidence counting, require the arrival of photons at different SNSPDs to be time-tagged using a precisely synchronized base clock. The number of input channels on these devices also continues to grow \cite{fleming2025high,oripov2023superconducting}. For these reasons, many SNSPDs are often housed inside the same cryostat, and then all connected to the same TDC (Fig. \ref{fig:detector_channels_match}). The resulting profusion of fiber-optic and electronic cables makes it difficult to identify which ports on the TDC are connected to which MEAS and BSA nodes.

We first propose a straightforward network protocol and two possible algorithms that use this protocol to automate the configuration of SNSPDs and TDCs.


\subsubsection{Automatic switch-node configuration}

Optical quantum network nodes may be divided up into three categories:

\begin{enumerate}
    \item Source nodes, such as EPPS, which produce light
    \item Detector nodes, such as MEAS and BSA, which consume light
    \item Transit nodes, such as optical switches (OSW) which pass through light
\end{enumerate}

It is this last category, transit nodes, and specifically OSW nodes, which transform a quantum optical system from a single, static link into a dynamic network. However, to state the obvious, switches are only useful when we know what nodes are attached to them. Without this information, we cannot construct a network topology model or determine how to route between a source and a detector.

However, compared to modern classical network switches, current optical switches are often simple pass-through devices, creating a lightpath between an input port and an output port with no knowledge of whether light is actually traveling along that path, let alone the ability to identify or produce classical optical signals that might help identify end nodes. Adding such hardware may also cause additional photonic loss, decreasing the overall efficiency of the system.

In the second part of the paper, we propose protocols and algorithms to automate the process of identifying what is connected to each port of all switches in a multi-switch optical quantum network. We consider the minimum additional hardware necessary to implement these tools.
\section{State of the art}
Following the proposal of unknown quantum state teleportation using the distribution of Bell pairs  and transferring the measurement results via classical channels some decades ago\cite{bennett1993teleporting}, many successful experiments have demonstrated the ability to generate, distribute, and transfer entanglement and quantum states \cite{bouwmeester1997experimental, ursin2007entanglement, pompili2021realization, krutyanskiy2023entanglement}. In the early evolution of entanglement-based quantum networks, Bell pairs were distributed between two quantum nodes. For several years, experiments focused on extending entanglement distribution distance, rather than on increasing the number of nodes. However, in recent years, there has been progress in the development of multi-node quantum networks\cite{Alshowkan22,Krutyanskiy2023ion,QUANTNET2024,sakuma2024optical}, bringing experiments closer to real-world quantum networks.


One of the first experimental systems for the management of a quantum network link using a link layer protocol is proposed in \cite{dahlberg2019link}. It mainly focuses on the monitoring and calibration of a single logical quantum links in terms of entanglement generation rate and performance metrics such as fidelity.

As in classical networks, which have multiple layers of communication and protocols, quantum networks also require a multi-layered communication framework with corresponding protocols \cite{pompili2021realization, delle2025operating, oka2016classical, matsuo2019quantum}. For example, from the early development of IEEE 802.3 \cite{IEEE802.3} standards to the present, Ethernet has evolved from 10base-T to the current standardization of Terabit Ethernet, defining various media access control (MAC) and physical layer operations. On top of these protocols, additional protocols are used to translate network addresses to MAC addresses, such as ARP \cite{plummer1982rfc0826} in IPv4 and NDP \cite{narten2007rfc} in IPv6. Furthermore, protocols such as LLDP \cite{IEEE802.1AB} can be used for network topology discovery. In this work, we focus on the lowest layers of communication for quantum networks, and explore potential quantum equivalents for MAC addresses and network address translation.

In addition to Ethernet-based networks, we can also compare optical quantum networks with all-optical, high-speed classical networks. In these optical networks, no optical to electrical signal conversion is done. Lambda (wavelength) switching and multiplexing are performed using optical cross connect (OXC) switches and reconfigurable optical add-drop multiplexers (ROADM). GMPLS is used as a generalized control plane to manage these networks \cite{rfc3472,rfc4206}. However, the single-mode nature of current optical quantum network testbeds makes it difficult to re-use these same techniques at the present time.

\section{Design Strategies} 
\subsection{Assumptions} 
This work is primarily concerned with the detection of \textit{channels}: the physical links between quantum network nodes. When designing protocols to address this need, we will consider the real configuration of current testbeds as well as future trends in their evolution.

For detailed analysis of different strategies, we divide channels into several categories. Table. \ref{tab:channel_types} describes these categories and their permutations.

\begin{table}[htbp]
    \centering
    \caption{Enumeration of all possible light paths between light source (S-type), transit (T-type), and detector (D-type) nodes. This table does not include repeater and router nodes used for long-distance network communication, but the same principles apply: light travels in one direction from S-type or T-type nodes, and terminates at D-type nodes.
    }
    \begin{tabular}{|c|c|c|c|c|c|}
        \hline
        & OSW & BSA & MEAS & EPPS & COMP \\
        \hline
        \hline
        OSW & $T \rightarrow T$ & $T \rightarrow D$ & $T \rightarrow D$ & N/A &  $T \rightarrow D$\\
        \hline
        BSA & N/A & N/A & N/A & N/A & N/A \\
        \hline
        MEAS & N/A  & N/A & N/A & N/A & N/A \\
        \hline
        EPPS & $S \rightarrow T$ & $S \rightarrow D$ & $S \rightarrow D$ & N/A & $S \rightarrow D$ \\
        \hline
        COMP & $S \rightarrow T$ & $S \rightarrow D$ & $S \rightarrow D$ & N/A & $S \rightarrow D$  \\
        \hline
    \end{tabular}

    \label{tab:channel_types}
\end{table}

During the development of the protocols, we have this set of assumptions: 
\begin{enumerate}
    \item The channel discovery protocols are intended for memoryless optical quantum networks, where flying qubits are sent from node to node as a stream of photons.
    \item The protocols are not dependent on any particular optical technology or property. They should work for any wavelength of light.
    \item The flow of light in the network is inherently one-way. Nodes have physically separate input and output ports. We assume that each node knows its own port configuration before the discovery protocols run.
    \item A channel's source node can control the emission of light at its output ports.
    \item Network nodes may be physically disjoint, necessitating the need for even lower-level protocols than normally found in classical networks. For example, single-photon detector and time-to-digital converter hardware is often shared between nodes. A protocol is presented in Section \ref{DetectorProto} to address this issue.
\end{enumerate}

\subsection{Approaches}
Focusing on different properties of components and devices in quantum networks can open up new paths for addressing configuration problems. The main relevant properties can be enumerated as follows:
\subsubsection{Detector data analysis}
    This approach makes use of the number of detection clicks and their start times. While it can be useful in certain parts of the network, it is not a complete solution. This is because some channels between T-type ports may also be part of a light path, which complicates the analysis.

\subsubsection{Controlling photon source devices}
    This approach involves stopping and starting the main lasers used as light sources to identify active channels between nodes in the network, for example, by controlling pulsed lasers. However, there are several challenges with this method. First, compatibility can be an issue, as this approach is often vendor-specific: some devices do not support automated control. Additionally, re-initializing a pulsed laser can be a difficult task. Finally, due to varying configurations, a pulsed laser may directly connect to multiple nodes, which can lead to the simultaneous activation of multiple light paths across the network.

\subsubsection{Monitoring channel endpoints}
    The third approach involves monitoring the presence of light in physical quantum channels. This is the same method often used by experimentalists for troubleshooting and verifying light paths during experiments. Typically, they manually block the light path or measure light power using a power meter sensor. Monitoring physical channels in free-space setups is straightforward. In fiber-based systems, certain optical modules allow for digitally controlling the attenuation, enabling control of the light inside the fiber. Some optical switches \cite{Polatis} also incorporate light sensing capabilities. Even in the case of less advanced networks, as we will see in Sec. \ref{Node-to-Node}, it is possible to monitor and discover all types of channels using simple tools.

In this work, we adopt the third approach: monitoring the light between nodes' input and output ports to identify the physical channels between them. Monitoring requires only the detection of light's presence or absence at a port. More complex techniques such as Hong-Ou-Mandel or tomography are not employed. This makes the method simple and broadly applicable to many different optical quantum network configurations.

\subsection{Quantum Network Discovery Flow} 
\begin{figure}
    \centering
    \includegraphics[width=1\linewidth]{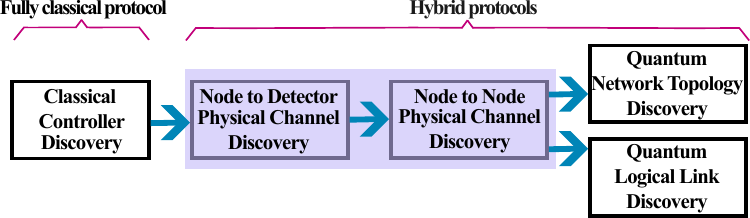}
    \caption{Quantum Network Discovery Flow. The highlighted sections show the scope of current work.}
    \label{fig:DiscoveyFLow}
\end{figure}

In quantum networks, the classical control plane and the quantum data plane are distinct and involve multiple layers. Therefore, the discovery of devices, nodes, and networks across different planes and layers requires specialized processes and protocols. As illustrated in Fig.\ref{fig:DiscoveyFLow}, the discovery protocol in this context can be divided into several stages.

Classical controller discovery is not the focus of this paper: depending on the network topology, it can be accomplished using any number of existing low-level classical protocols such as UPnP or mDNS\cite{cheshire2013multicast} or higher-level service discovery systems. However, in order to discover quantum channels, node controllers must coordinate actions in both the classical and quantum networks. Once the channels are identified, each node will be fully aware of its neighbors. This in turn will enable the development of higher-layer protocols for the discovery of quantum network topology and even logical quantum links, such as the MIM (Memory-Interference-Memory), MSM (Memory-Source-Memory), and MM (Memory-Memory) schemes\cite{soon2024implementation,jones2016design}.

\section{Problem 1: Identifying single-photon detector channels} \label{DetectorProto}
First, we propose a protocol and algorithms to associate superconducting nanowire single-photon detectors (SNSPDs) with their corresponding network nodes: specifically, measurement (MEAS) and Bell-state analyzer (BSA) nodes.

The configuration of every layer of the quantum networking laboratory stack, from the bottom up, must be automated in order to enable larger and more complex experiments. In optical networks, one of the lowest layers of connectivity is that between network nodes and SNSPDs. The number of detectors grows linearly in proportion to the number of nodes in the network; as each measurement or BSA node requires their own detectors, even a simple two-end-node system with a BSA in the middle will use at least four detectors. This creates a profusion of wiring that must be manually traced and configured before starting system tests.

Each SNSPD is connected to a time-to-digital converter (TDC) input channel. Each time a photon is detected by the SNSPD, it emits an electrical pulse that is received and time-tagged by the TDC.

For practical reasons, the TDC is connected to a single classical computer which receives data from all connected channels: this computer runs a network service to which other nodes can make requests. We consider this \textit{TDC service} to be the source of truth for the correspondence between TDC channels and nodes. Other network nodes will query the TDC service to learn how they are connected to the TDC.

Aside from the TDC service, each node is controlled by a separate network service. These services are aware of the network address of the TDC service, and are considered the source of truth for the state of the measurement and BSA nodes. Importantly, these \textit{end-node services} have the ability to start and stop the flow of laser light through the node to the SNSPD.

At the beginning of an experiment, the TDC service is in an \textit{unconfigured} state, meaning that the service knows how many network nodes are connected to it, but does not know which TDC channels correspond to which network nodes. The network cannot begin processing requests until the TDC service has reached a \textit{configured} state. To reach this state, the endpoint and BSA services must make identification requests to the TDC service. These requests may be serviced using one of the following algorithms.

In the subsections below, ``node'' is used as a shorthand for a hypothetical network node which incorporates exactly one SNSPD. In reality, most network nodes will have more than one detector; the protocol definition that follows description of the algorithm incorporates that reality.

Also implicit in these proposals is that unpolarized light is always flowing into the nodes. In a real network testbed, light flows into these nodes through an optical switch, and many of the available light sources may be polarized.

Finally, we assume low-latency transmission of classical messages, with delivery and processing taking no more than a few hundred milliseconds.

\subsection{Serial identification of TDC channels} 
The simplest way to move from an unconfigured state to a configured state is for each of the nodes to identify itself to the TDC service serially.

At system power-on, all nodes set themselves to an off position. In other words, at this point, although light is flowing into all of the nodes, no light is flowing from the nodes to the TDC. Once the TDC service validates that there is no light flowing to any of its channels, it begins to allow identification requests from nodes. The TDC service takes the first identification request it receives and sends a reply allowing the requesting node to proceed by turning itself on. All other requests receive a reply asking them to wait and retry after a specified period of time.

When the first requesting node turns itself on, the TDC service is able to identify which channel the node is connected to by simply looking for an unconfigured channel that has light arriving to it. At this point, the TDC service sends another message to the requesting node, informing it that configuration of the requested channel is complete. At the same time, the TDC marks that channel as configured. The TDC service then repeats the same process until all channels have been identified and marked as configured.

The approximate runtime is $d \times t$, where $d$ is the total number of detectors in the system and $t$ is the time it takes for a device to transition from off to on: note, however, that the device may not need to be completely on for the TDC to register the change, which could result in significant performance improvements. This approach may have some performance advantages in systems that are slow to stop or start the flow of light. Using free-space blocking techniques, for example, by closing a shutter, it may take more than one second for a node to fully transition from an off to on state, or the reverse. Since each node remains in an on state after identification, the TDC service can immediately transition from an unconfigured state to a configured state as soon as all channels have been identified.

\subsection{Parallel Identification} 

A more complex approach to identifying channels is to allow more than one node at a time to identify itself. In this model, all nodes concurrently send light patterns that uniquely identify each node.

As with serial identification, all nodes are assumed to start in an off position. Once the TDC service validates that there is no light flowing to any of its channels, it begins to allow identification requests from nodes.

When the TDC service receives its first identification request, it responds as in serial identification, sending a reply allowing the requesting node to proceed. However, unlike in serial identification, the requester is asked to modulate its light in a pattern specified by the TDC service.

Subsequent requests received while the first request is being fulfilled are not delayed as in serial identification, but rather immediately accepted. For each request, the TDC service assigns a different light modulation pattern.

The TDC service watches all available channels. Each time the TDC service identifies a known pattern, the service informs the requester that identification was successful and marks the associated channel as configured. The requester can then turn itself on and await normal network operations. When the service has verified that all channels are on, it transitions from the unconfigured state to the configured state.

There are many possible ways to send a light pattern to the detector; the approach used will depend on the available technology. Choosing from available, well-established protocols for sending data over optical fiber is thus outside the scope of this paper. We will instead proceed with a simple, generic example for analysis.

In this example, a bit is represented by a node turning itself on or off for a duration specified by the TDC service.

TDCs typically have 255 or fewer channels. Consequently, it should be possible to identify all connected channels simultaneously by sending an 8-bit pattern.

It must be possible for the TDC to identify the beginning and end of this 8-bit pattern. One simple way to do this could be to send the header \verb|0111111110| before the 8-bit pattern, and using the 8-bit pattern to represent values up to 254, thus reserving 255 for the header value.

Assuming that all nodes request identification simultaneously, the approximate runtime is $rt(2\lceil\log_2{(d+1)}\rceil + 2)$, where $d$ is the total number of detectors in the system, $r$ is the number of times the pattern must be repeated, and $t$ is the duration of time required to represent each bit.

Even from this simple example, we can predict that for non-trivial systems parallel identification should be significantly faster than serial. However, real-world designs must take into account further considerations, such as software complexity, the difficulty of delivering light to all nodes at once, and the relative speed (or lack thereof) of signaling hardware which may make the serial option more appealing.

\subsection{Protocol specification} 
The same protocol can be used to implement both of these approaches. In fact, it is not necessary for network nodes to know which strategy the TDC service is using. The messages described here are abstract; they may be implemented using any RPC architecture. They form part of the TDC service API. Other parts of the API may be concerned with network operation after the service has transitioned to a configured state, and are not described here.

\subsubsection{Starting identification}

\begin{itemize}
    \item \verb|ID_REQ|: The identification process is initiated by the node sending a \verb|ID_REQ| message to the TDC service. The message contains the following information:
    \begin{itemize}
        \item \verb|NODE_ID|: An identifier for the node. This must be unique among all nodes connected to the TDC service. In our current model, this ID has been pre-registered with the TDC service. In the future, sending this request may cause the TDC to dynamically register a new node.
        \item \verb|DETECTOR_ID|: An identifier for the specific detector associated with this node to identify. Must be unique within the node.
    \end{itemize}
\end{itemize}

The response depends on the state of the TDC service, and could take one of three forms:
\begin{itemize}
    \item  \verb|ID_START|: Start the identification process immediately.
    \begin{itemize}
        \item \verb|DURATION|: If 0, indicates that the node should turn on and remain on. If greater than 0, indicates how long the node should remain on before turning off again.
        \item \verb|PATTERN|: A string of arbitrary length consisting of the characters 0 (off) and 1 (on), indicating a light pattern the node should repeat until it is told to stop. If \verb|DURATION| is 0, this field should be omitted.
    \end{itemize}
    \item \verb|ID_RETRY|: The identification process is not available right now, but will be at a later time. The requester should send another \verb|ID_REQ| message later.
    \begin{itemize}
        \item \verb|WAIT|: The duration that the requester should wait before retrying.
    \end{itemize}
    \item \verb|ID_ERROR|: The identification process is not available for some reason.
    \begin{itemize}
        \item \verb|TYPE|: The type of error, if known. For instance, in the current implementation, if the TDC service is already in a configured state, it cannot fulfill \verb|ID_| requests.
    \end{itemize}
\end{itemize}

\subsubsection{Completing identification}
After sending an \verb|ID_START| command, the TDC service is responsible for notifying the requester upon completion of the identification process. It returns one of two types of messages: 
\begin{itemize}
    \item \verb|ID_COMPLETE|: The TDC service has successfully associated a channel with an \verb|ID_REQ| request. After receiving this message, the node should turn the flow of laser light to this detector on and allow it to remain on.
    \begin{itemize}
        \item \verb|NODE_ID|: The node identifier sent in the corresponding \verb|ID_REQ| request
        \item \verb|DETECTOR_ID|: The detector identifier sent in the corresponding \verb|ID_REQ| request.
        \item \verb|CHAN_ID|: The TDC channel ID.
    \end{itemize}
    \item \verb|ID_ERROR| messages may also be sent by the TDC service at this point, for example, if a timeout occurs.
\end{itemize}

\subsubsection{Lookup}
Any node participating in the network may send a lookup request to the TDC service at any time to learn the channel ID associated with any node-detector pair. 
\begin{itemize}
    \item \verb|ID_LOOKUP_REQ|: Look up the corresponding channel ID, if known, for a node-detector pair. The message includes \verb|NODE_ID| and \verb|DETECTOR_ID|. 
\end{itemize}

The TDC will reply with one of three messages:
\begin{itemize}
    \item \verb|ID_LOOKUP_CONF|: The corresponding channel has been identified. Included data fields (\verb|NODE_ID|, \verb|DETECTOR_ID|, \verb|CHAN_ID|). 
    \item \verb|ID_LOOKUP_UNCONF|: The corresponding channel has not yet been identified, but the node and detector are known to the TDC. It contains  \verb|NODE_ID| and \verb|DETECTOR_ID|.
    \item \verb|ID_LOOKUP_ERROR|: The lookup failed for some reason, for instance, if the node is not known to the TDC.
\end{itemize}

\subsubsection{Configuration Status}
Any node participating in the network may send a status request to the TDC at any time. Nodes may poll the TDC service to determine when it is appropriate for them to start making network requests.

\begin{itemize}
    \item \verb|ID_STATUS_REQ|: Request the configuration state of the TDC service.
    \item \verb|ID_STATUS|
    \begin{itemize}
        \item \verb|STATUS|: One of \verb|UNCONFIGURED| or \verb|CONFIGURED|. \verb|CONFIGURED| indicates that the TDC service is ready for network operations. In the current model, \verb|CONFIGURED| also implies that the TDC service is no longer accepting \verb|ID_REQ| requests, but this is not necessarily true of future, more dynamic implementations.
    \end{itemize}
    \item \verb|ID_STATUS_ERROR|
 \end{itemize}

\subsubsection{Flow}
\begin{figure}
    \centering
    \includegraphics[width=0.8\linewidth]{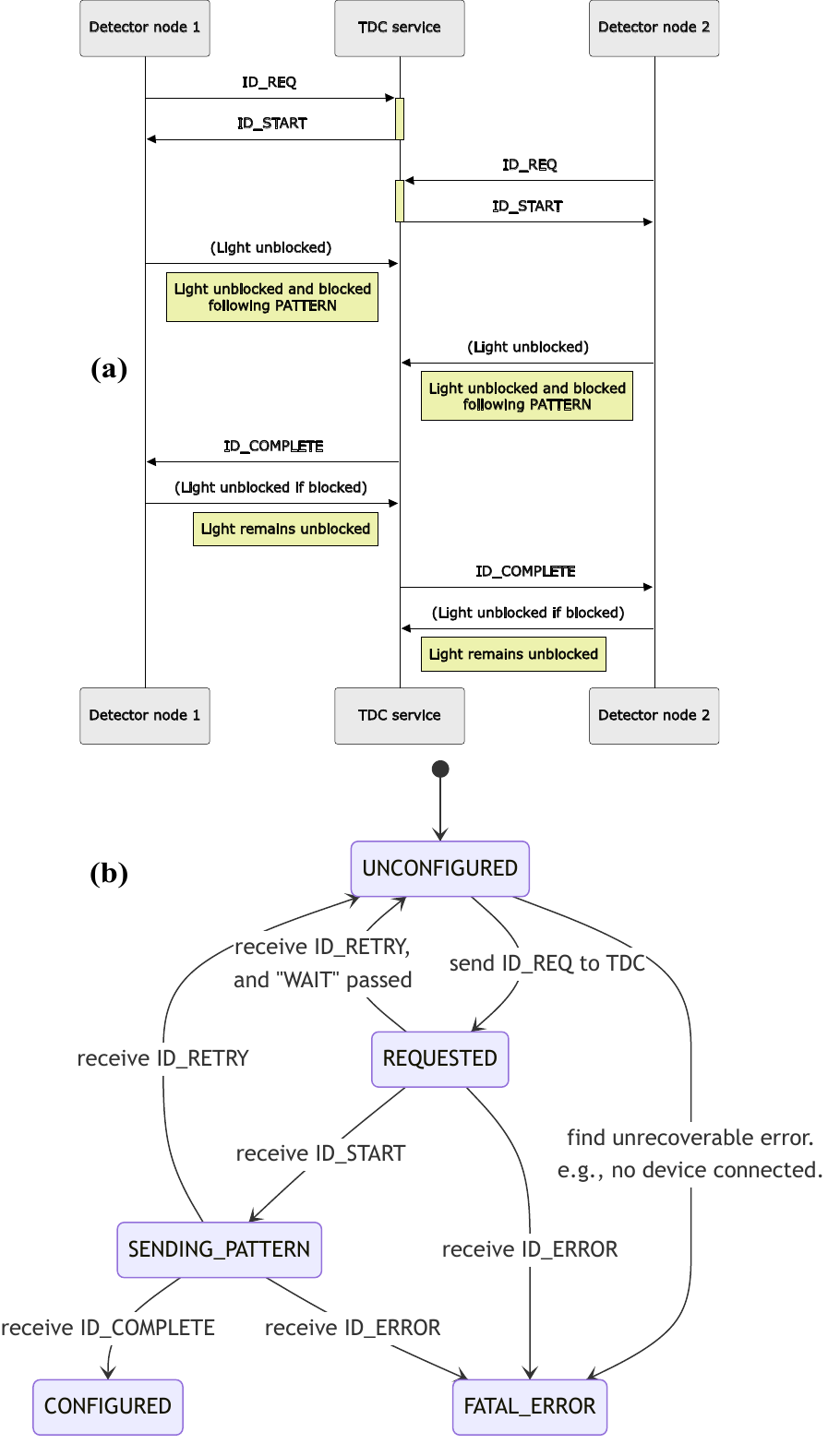}
    \caption{(a) A flow for parallel identification of detector channels for two end-points. (b) State diagram of the parallel identification approach in automatic Node-to-TDC configuration. The blue lozenges are states, and the labels on arrows  indicate the event that triggers a particular transition.}
    \label{fig:parallel-seq-diagram}
\end{figure}

Fig. \ref{fig:parallel-seq-diagram} illustrates the message flow for the parallel configuration of two endpoints along with the corresponding state diagram.

\subsection{Expansion to multiple TDCs}
In the future, larger networks will likely employ multiple, geographically separated TDCs. However, as there is no need to share TDC channel ID information outside the scope of the nodes connected to a single TDC, this protocol should work equally well in such configurations. From a logical standpoint, a network node consists of its own optical hardware combined with its associated TDC channels. Although defining the service model for network nodes is beyond the scope of this paper, it is clear that the association between node detector ID and TDC channel ID can be encapsulated in the nodes' own service instance, and does not need to be shared with other nodes.

\subsection{Security and robustness considerations}
Given that the design already presupposes that the TDC is aware of exactly how many nodes are connected to it and how to identify those nodes, it is implicit that all nodes in the network are trusted and an operator is available on-site to troubleshoot any issues. To that end, the outline above has largely elided important considerations such as authentication, timeouts, and rate limiting that may be added later.

In the planned implementation, the TDC will transition from an unconfigured to a configured state when all channels have been identified. This allows badly-behaved, broken, or malicious nodes to deny service to all nodes by preventing the TDC service from transitioning to the configured state. This is not an inherent limitation in the protocol, and, in the future, the distinction between unconfigured and configured may be different; network operations may begin before channel identification operations have completed. Technical limitations in current-generation TDCs and software make such a system difficult to implement at this time.

\section{Problem 2: Identifying node-to-node channels} \label{Node-to-Node}
The second problem we approach is the identification of lightpaths or channels between network nodes.

Based on the current capabilities of quantum network hardware, we consider two design models: (1) single-pulse light coding coupled with a publish-subscribe–based mechanism, (2) pattern-encoded light coupled with an optical address resolution lookup mechanism.

The first model, with its orchestration capabilities, is best suited for structured networks, such as groups of switches in data centers. The second model is more appropriate for general local area networks. It is also more compatible with early-stage, limited-capability quantum network hardware.

\subsection{Single on-off pulse coding with pub-sub messaging}


In this conceptual model, all active channels within the optical domain can be incrementally identified by selectively unblocking light from node output ports for specific durations of time, while simultaneously monitoring potentially corresponding input ports at the opposite ends of the physical channels. The dynamic behavior of the physical layer is communicated through a classical publish/subscribe messaging system, acting as an event bus accessible to all classical controllers within the optical domain.

We assume that each node is aware of the types of its ports (input/output), and consequently, the Source, Transit and Detector port type category as shown in Table \ref{tab:channel_types}. Further, each S- and T-type node can block and unblock the light emitted from its output ports. We also assume that, before reaching a configured state, every D-type and T-type input port can measure incoming light. For D-type input ports, the node's built-in detectors, such as SNSPDs, can be used for measurement. For T-type input ports (which exist only on OSW nodes in the optical domain), a general low-cost light power sensor is sufficient to detect light and measure pulse duration. As we will see in the next section, this last assumption does not need to be strictly enforced. It can be relaxed by instead placing an auxiliary detector or power sensor on one of the predefined output ports of the switch. 

The protocol begins with S-type nodes. These nodes will choose a random start time and duration for which to unblock their source light, and then publish a message to the $channel/active$ message bus topic. At the same time, these nodes subscribe to the $channel/detect$ topic.

All other nodes in the optical domain with input ports (T-type or D-type) subscribe to the $channel/active$ topic and at the same time watch for incoming light on their unconfigured input ports. If light is detected, the node checks the pulse duration and publishes a message to the $channel/detect$ topic.

After completing light pulse generation and monitoring along their lightpaths and exchanging pub/sub messages in the classical control plane, the nodes enter the matching phase. In this phase, each node processes the received messages based on activation time, detection time, and pulse duration. By filtering out unrelated events, nodes can construct a candidate list from the remaining messages. This list contains possible peers that may have matched during the channel activation and detection phases, typically when two or more pulse generation events occur simultaneously and with the same duration. Next, nodes verify the connectivity of the corresponding channel. This verification may involve repeating the pub/sub messaging using the same topic as in the previous step. Repeating this process helps reduce conflicts and increase the probability of selecting the correct peer from the candidate list. 

\begin{figure}
    \centering
    \includegraphics[width=1\linewidth]{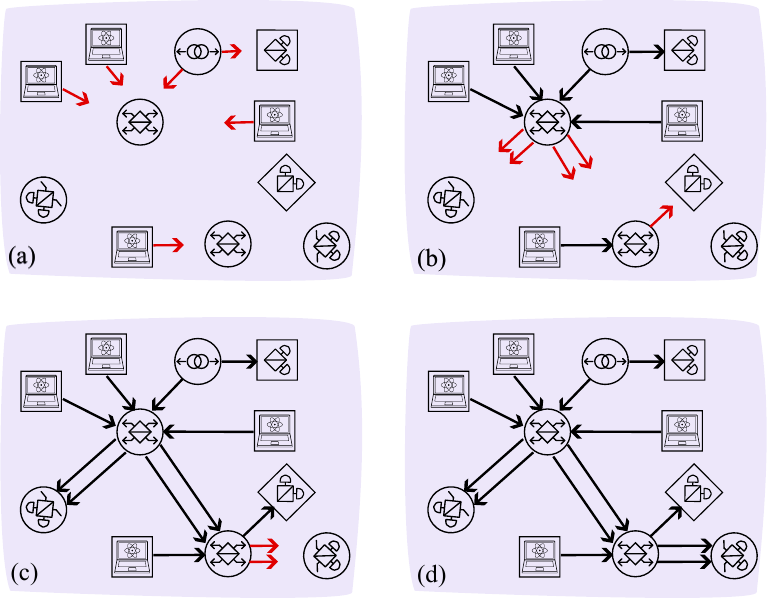}
    \caption{The pub/sub-based protocol for identifying inter-node channels. (a) The S-type nodes randomly activate their output ports at different times and transmit light for random durations. In the control plane, these nodes publish port activation messages. Meanwhile, all input ports are in listening mode, having subscribed to channel activation messages. Nodes that detect incoming light pulses publish channel detection messages. Each node then attempts to correlate activity and generate a candidate list for each port. Direct negotiation is used to verify the actual channel connectivity between nodes. (b), (c) The protocol proceeds incrementally, identifying additional channels by incorporating T-type nodes. (d) Eventually, all channels have been identified.}
    \label{fig:Single_on_off_pulse_coding}
\end{figure}

After discovering the channels for S-type output ports, as shown in Fig. \ref{fig:Single_on_off_pulse_coding}, light with a known source can be forwarded to other channels. This process continues with T-type ports, which correspond to the switch ports. The procedure is repeated at each node in the domain until all active channels are discovered for every port on each node.


Although the basic assumption is that the input ports of nodes with transit ports (i.e., switches) are capable of monitoring light power on their input ports, this requirement can be relaxed by configuring the switch to include an auxiliary detector on one of its output ports, as shown in Fig. \ref{fig:AuxiliarySensor}. 

\subsubsection{Protocol}
The processes necessary for monitoring input ports and activating output ports are shown in Algorithm \ref{PUBSUBMonitor} and Algorithm \ref{PUBSUBActivate} respectively. 

\subsubsection{Message Semantics}
Messages sent to topic $channel/activate$ contain an activation time $T_{i}$ representing the time when the output port is unblocked, as well as duration time $D_{i}$, indicating how long the port will continue to emit light. Properties \verb|NODE_ID| and \verb|PORT_ID| identify the message sender.

Messages sent to topic $channel/detect$ contain a detection time $T_{i}$, representing the time when the input port found an activation pulse, as well as $D_{i}$, the duration of the detected pulse.

To set the activation and duration times, each node starts by randomly selecting values from the intervals $[0,T_{init}]$ and $[0,D_{init}]$, respectively. If a node detects that two different nodes have published messages with the same duration 
$D_i$, it increases both its activation time and duration parameters. During the detection phase, if two published messages have the same detection time and duration, the node doubles its activation and duration times. However, if a conflict is detected during the verification phase, the node interprets this coincidence as a sign of congestion and halts channel activation by setting its activation time to the maximum value. After a timeout period, the node resumes listening to published messages, and if no further coincidences are found, it gradually decreases the activation time. 

\begin{algorithm}[htb]
\DontPrintSemicolon
\SetKwInOut{Input}{Input}\SetKwInOut{Output}{Output}
PROCESS PortMonitor() \\
\While{there exists an input port $P_i$ such that $P_i.state = \texttt{UNCONFIGURED}$}{
  \ForEach{$P_i$ in InputPorts}{
    subscribe to topic \texttt{"channel/active"}\;
    
    \uIf{message $m$ is received from \texttt{"channel/active"}}{
      publish a message with topic \texttt{"channel/detect"}\;
      candidate $\leftarrow$ extract(filter($m$))\;
      
      \If{$candidate \neq \emptyset$}{
        add candidate to $CandidateList[P_i]$\;
        
        \If{verify(candidate)}{
          $P_i.state \leftarrow \texttt{CONFIGURED}$\;
        }
        \Else{remove candidate from $CandidateList[P_i]$}
      }
    }
  }
}
\caption{Single-Pulse Coding Pub/Sub Protocol- PortMonitor Process for input ports \label{PUBSUBMonitor}}
\end{algorithm}

\begin{algorithm}[htb]
\DontPrintSemicolon
\SetKwInOut{Input}{Input}\SetKwInOut{Output}{Output}
PROCESS PortActivate() \\
\OPL $\leftarrow$ list of all UNCONFIGURED output ports $P_i$\;
\While{\OPL $\neq \emptyset$}{
  \ForEach{$P_i \in$ \OPL}{
    $T_i \leftarrow$ select random activation time\;
    $D_i \leftarrow$ select random duration\;
    activate $P_i$ at $T_i$ for duration $D_i$ in physical channel\;

    publish message to topic \texttt{"channel/active"}\;
    subscribe to topic \texttt{"channel/detect"}\;

    \uIf{message $m$ is received from \texttt{"channel/detect"}}{
      $candidate \leftarrow \texttt{extract(filter(}m\texttt{))}$\;
      \If{$candidate \neq \emptyset$}{
        \If{verify($candidate$)}{
          $P_i.state \leftarrow \texttt{CONFIGURED}$\;
          remove $P_i$ from \OPL
        }
      }
    }
  }
}

\caption{Single-Pulse Coding Pub/Sub Protocol- PortActivate Process for output ports \label{PUBSUBActivate}}
\end{algorithm}

\subsection{Pattern decoding}
This model does not assume that a centralized message broker is available for pub/sub services. Instead, for each channel, the controller of the two nodes on the source side and sink side of the channel should find each other by interpreting the behavior of light on both ends of the channel. To accomplish that, the source node of the channel encodes a pattern by sequentially blocking and unblocking its light source. The node on the sink side of the channel observes incoming light and decodes the pattern. This pattern should be unique inside the network namespace. The value encoded in the pattern is conceptually similar to a classical network adapter's MAC address.

After decoding the pattern, the receiving node can query the network to resolve the sending node's IP address and \verb|NODE_ID|. This is much like using reverse address resolution protocol to find the IP address of a node from the MAC address in the network.  

\begin{figure}
    \centering
    \includegraphics[width=1\linewidth]{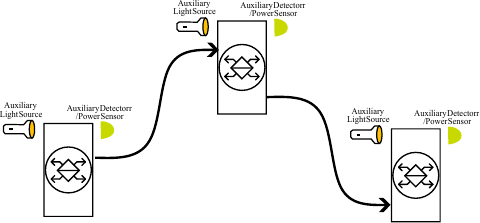}
    \caption{Instead of a light sensor at every input port of the switch, a single auxiliary sensor can be attached to an output port on each switch. However, in addition to the loss of a useful output port on each switch, identification operations must proceed serially rather than in parallel. Adding a controllable auxiliary light source to each switch could also simplify the process of identifying D-type ports. Such a scheme can eliminate the uncertainty caused by chained switches during node identification.}
    \label{fig:AuxiliarySensor}
\end{figure}

As in the first model, we assume that each optical switch input port has a built-in light sensor. However, as shown in Fig. \ref{fig:AuxiliarySensor}, we can instead assume that one input port of the switch is connected to a light source, and one output port is connected to a light detector. In this setup, a single light source and a single detector are sufficient for each switch to identify all of its ports.

\subsubsection{Protocol Specification}
Algorithm \ref{ClientServer} illustrates the protocol structure.
In this approach, all S-type and T-type nodes can announce their \verb|NODE_ID| along with a list of \verb|PORT_ID| and \verb|PATTERN| pairs. The announcements are collected into a lookup table. The mechanism used to make announcements and distribute the lookup table is outside the scope of this paper; DNS is one possibility in a distributed system.

S-type nodes are the first to make announcements; they may all announce in parallel. After announcement, each node begins encoding its \verb|PATTERN| in its output light. Meanwhile, each T-type switch node, if equipped with an auxiliary light source, can sequentially identify active channels for its T-type output ports. For each T-type output port, the node transmits a pattern and waits to receive a notification from a D-type or T-type input port of another node.

On the receiving side, nodes with D-type and T-type input ports listen for incoming light signals. In the case where a switch has only one auxiliary detector or power meter and multiple T-type input ports, the detector is assigned to each input port sequentially. The input ports watch for light patterns; once detected, the node decodes the pattern and finds the \verb|NODE_ID| and \verb|PORT_ID| via the optical address lookup table. After resolving \verb|NODE_ID|, the sink node notifies the source node. After confirmation by both sides, the corresponding \verb|NODE_ID| and \verb|PORT_ID| are added to the neighbor table. Each entry in the neighbor table has a validation time and can be refreshed. Upon expiration, the entry is removed, and the nodes need to run the protocol again.


\begin{algorithm}
\DontPrintSemicolon
\SetKwInOut{Input}{Input}\SetKwInOut{Output}{Output}
\KwIn{An optical domain with $N$ nodes. Each node has a list of input ports and output ports with optical pattern IDs related to its own}
\KwOut{Updated neighbor table contains the node\_id and port\_id of neighbors}
PROCESS Pattern\_Decoding\_Protocol() \\
\ForEach{node $node\_id$ in the optical domain}{
     Announce $(node\_id, port\_id, optical\_pattern\_id)$\;
}
\ForEach{output port $P_i$}{
    \If{$P_i$ is T-type}{
        Reserve a light source\;
    }
    \Repeat{the sink node is confirmed}{
        Generate $optical\_pattern\_id$ in physical channel\;
        Notification is received from a sink node\;
        Confirm the sink node\;
    }
    Add $(sink\_node\_id, sink\_port\_id)$ to the neighbor table\;
}
\ForEach{input port $P_j$}{
    \If{$P_j$ is T-type} {
        Reserve a light source\;
    }
    \Repeat{The source node is confirmed}{
    Listen for incoming optical patterns\;
        Upon receiving pattern $OP$ 
        {
            \quad $(source\_node\_id, source\_port\_id) \leftarrow$ lookup in pattern table using $OP$\;
            \quad Notify $source\_node\_id$\;
            \quad Wait for confirmation\;
        }
    }
    Add $(source\_node\_id, source\_port\_id, OP)$ to the neighbor table\;
}
\caption{Pattern-based channel identification \label{ClientServer}}
\end{algorithm}
\section{Evaluation} \label{Evaluation}
Finally, we analyze the protocols with respect to runtime and errors. Quantum networks are highly time-sensitive and contain many potential sources of delay\cite{timingregime2024}. Here, we consider the delay caused by the time required to generate and receive light pulses, $t_{act}$. We assume that the light propagation time is negligible at the scale of a data center quantum network. 

\begin{figure}
    \centering    \includegraphics[width=1\linewidth]{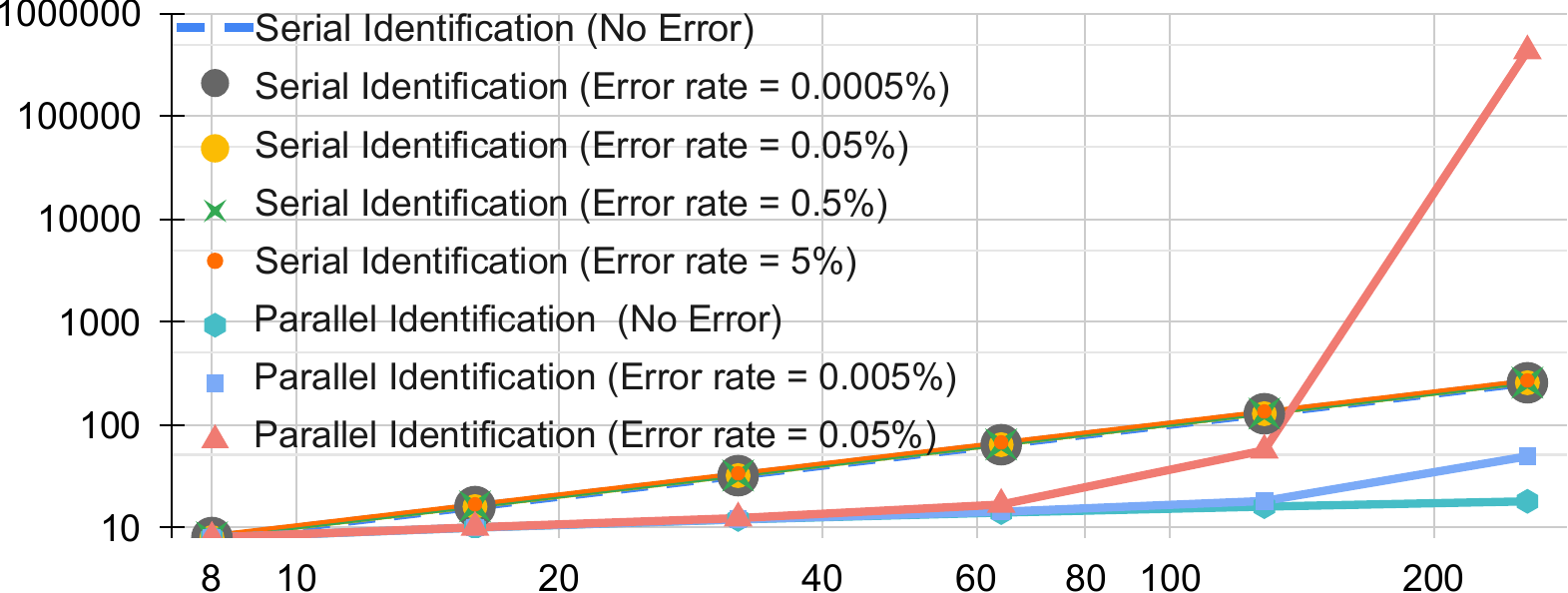}
    \caption{The estimated runtime (scale $t_{act}$) of node to TDC protocols given different geometrically distributed bit-level error rates, up to 256 detectors.}
    \label{fig:Eval-detector-channel-discovery}
\end{figure}

For node-to-TDC protocols, we consider light pattern encoding errors. Estimated runtimes for different conditions are shown in Fig. \ref{fig:Eval-detector-channel-discovery}. In the serial protocol, the time is linear with respect to the number of ports, even as error rates increase. However, for the parallel protocol, higher error rates quickly erase the protocol's lograrithmic advantage.

For the evaluation of node-to-node channel matching protocols, Table \ref{tab:proto_compare} provides a qualitative high-level  estimation of two protocols. 
\begin{table}[htbp]
    \centering
    \caption{Comparison of Node-to-Node protocols
    }
    \begin{tabular}{|c|c|c|}
        \hline
         & Single pulse pub/sub & Pattern decoding  \\
        \hline
        \hline
        Discovery time & Unpredictable & Predictable \\
        \hline
        Coding Complexity & Low (single light pulse) & Moderate \\ 
        \hline
        Message volume & High &  Low \\ 
        \hline
    \end{tabular}

    \label{tab:proto_compare}
\end{table}

\begin{figure}
    \centering    \includegraphics[width=1\linewidth]{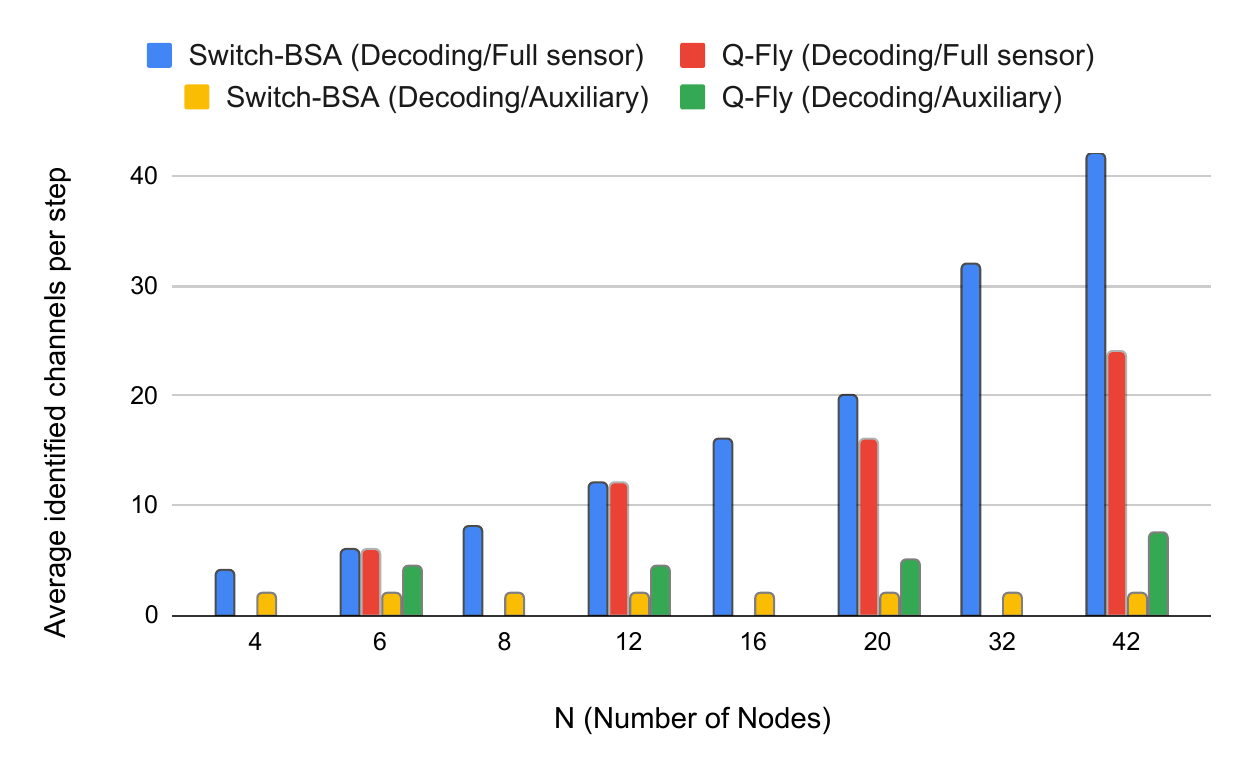}
    \caption{Average number of channels identified at each step of Pattern decoding based approach}
    \label{fig:Eval-channel-count}
\end{figure}

Further, Fig.~\ref{fig:Eval-channel-count} shows the estimated average number of channels identified at each step of the pattern decoding-based approach, using either fully monitored channels or auxiliary sensors, for two different topologies: a switch-BSA pool configuration~\cite{Koyama2024} and a small-scale Q-Fly with DPHD configuration~\cite{sakuma2024optical}, both with $N$ nodes. 


\section{Conclusion and Future Work}
In this work, we addressed challenges of automatic configuration in real quantum network testbeds, with an eye towards the implementation of fully automated optical domains at the data center scale. Specifically, we tackled two key problems: (1) automatic node-to-TDC configuration and (2) automatic switch-to-node configuration. To address the first, we proposed serial and parallel identification protocols. For the second, we introduced single on-off pulse coding with pub/sub messaging, as well as pattern-based encoding with optical address resolution. These solutions are applicable to a diverse variety of optical quantum network testbeds, without regard to implementation details such as light wavelength. For future work, we aim to deploy the protocols on a real-world quantum network testbed and evaluate results based on the real conditions and the gathered data.
\section*{Acknowledgments}
The authors would like to thank Hiroyuki Ohno, Michal Hajdu\v{s}ek, Naphan Benchasattabuse and the members of Nagayama Moonshot Project for their fruitful discussions and technical advice; Kent Oonishi, Kaori Nogata, Kaori Sugihara, Saori Sato, junsec, and Keio SFC Media Center for their help and support; and the many members of Keio SFC, AQUA, and WIDE for collaboration and support. ChatGPT and Overleaf's language tools were used to correct typographical and grammatical errors in some sections.  

\bibliographystyle{IEEEtran-new.bst}
\bibliography{IEEEabrv, bibfile}

\end{document}